\documentstyle[11pt,cite,epsf]{article}
\newcommand{\be}{\begin{eqnarray}}

\newcommand{\ee}{\end{eqnarray}}
\def\medfigsize{9 truecm}
\def\medcapsize{4.8 truecm}

\title{
	\begin{flushright}
	{\normalsize TPI--MINN--95--02/T \\
	NUC--MINN--95--8/T \\
	HEP--MINN--95--1327 \\
	hep-ph/9502289\\
	February 1995 \\}
	\end{flushright}
\bf Gluon Production from Non-Abelian Weizs\"acker-Williams Fields
in Nucleus-Nucleus Collisions
}
\author{
        Alex Kovner, Larry McLerran, and Heribert Weigert \\
	{\small\it School of Physics and Astronomy,
	University of Minnesota, Minneapolis, MN 55455}
	}

\date{}

\begin{document}

\maketitle

\begin{center}
{\bf Abstract}\\
\end{center}
We consider the collisions of large nuclei using the theory of
McLerran and Venugopalan.  The two nuclei are ultra-relativistic and
sources of non-abelian Weizs\"acker-Williams fields.  These sources
are in the end averaged over all color orientations locally with a
Gaussian weight.  We show that there is a solution of the equations of
motion for the two nucleus scattering problem where the fields are
time and rapidity independent before the collision.  After the
collision the solution depends on proper time, but is independent of
rapidity.  We show how to extract the produced gluons from the
classical evolution of the fields.

\vfill \eject

\section{Introduction}

Nucleus-nucleus collisions at ultra-relativistic energy have long been
recognized as an environment where hot dense matter is
formed \cite{bj,akm,bj1}. It has been conjectured that in such an
environment one might produce and experimentally study a quark-gluon
plasma \cite{review}. Theoretical studies of quark-gluon plasma
formation have typically assumed some initial conditions at some time
after the collision was initiated, and then evolved the matter
distributions forwards in time according to the equations of perfect
fluid hydrodynamics \cite{bj1,baym}.

While such an approach may work well for the late stages of the
collision when the particles are not so energetic, it does not work
well for the earliest stages of the collision.  In the earliest
stages, the quarks and gluons emerge from their quantum mechanical
wavefunction and cannot be described as a perfect fluid until at least
enough time has passed for there to be scattering.

In the earliest stages of the collision, the quark and gluon
interactions should be most energetic.  Such scatterings are therefore
more easy to experimentally probe as they presumably induce hard
experimental signatures which are more easily disentangled from
backgrounds due to soft final state processes.  During the
hydrodynamic expansion, typically the scale of energy in the
interaction is softer and more difficult to disentangle from
backgrounds.

There has been recent progress in attempting to describe the early
evolution of matter produced in nuclear collisions \cite{geiger}. In
the parton cascade model of Geiger and M\"uller, one takes the
experimentally measured distribution functions for quarks and gluons
and assumes that they may be treated as an incoherent beam of
particles arising from each nucleus.  The scattering of partons from
partons is computed making reasonable assumptions about quantum
coherence and time dilation effects.  The system is thereby evolved
from very early times in the collision until a later time when
hydrodynamics may be applicable.  In such a theory, the hard
scattering signals are computed and may be compared with experiment.

The parton cascade model while elegant and well motivated, in our
opinion still lacks some theoretical underpinning.  In particular, the
issue of quantum coherence in the initial state is treated
phenomenologically, and needs a deeper understanding.  This problem
has at least two important aspects.

The first and most glaring problem is that the partons arise from a
quantum mechanical state.  In such a state the uncertainty in
momentum, $\Delta p$, times the uncertainty in position, $\Delta x$,
is close to saturated,
\begin{eqnarray}
	\Delta p \Delta x \sim 1
\end{eqnarray}
For example, in the longitudinal momentum distribution of partons, the
wee partons have a longitudinal momentum of order of $100$'s of $MeV$.
This corresponds to a longitudinal size of order of fractions of a
Fermi.  On the other hand, in the parton cascade one assumes knowledge
of both the position and momentum of the partons, since the partons
are described by classical phase space distribution functions.  While
this should be true later in the collision as the scale of spatial
gradients becomes larger, early in the collision it is most certainly
violated.

Although in the parton cascade, the assumptions on the initial
distributions are plausible, they can at best give a qualitative
agreement with precise results which include the effects of coherence,
and at worst totally ignore some classes of interference phenomena.
For example, one obvious problem is that for a single nucleus, the
partons will spread out since they are an incoherent distribution of
partons with different momentum.  After some time, one therefore no
longer has a spatially compact nucleus.

Another class of phenomena which is not fully treated in the parton
cascade model is the problem of coherent addition of the color charges
of quarks and gluons.  Such coherent addition is for example
responsible for Debye screening, and presumably magnetic screening,
which will serve as a cutoff for divergent transport cross sections in
parton-parton processes.  In the parton cascade, a low momentum cutoff
is introduced by hand, and of course results for many processes depend
upon this cutoff.

While the parton cascade may lack precision in many detailed
computations, it nevertheless is outstanding for its qualitative
predictions.  We nevertheless would like to put this model onto firmer
foundations, and understand clearly its limits of applicability.

To begin to tackle this problem, one must understand at least some
aspects of the quantum mechanical wavefunction of the quarks and
gluons in the nuclear wavefunction. In the past, one rarely considered
the nuclear wavefunction, and the structure functions for a nucleus
were taken as a given quantity. There was no constructive description
of how such structure functions arise.

Recent work by McLerran and Venugopalan has given rise to a picture of
how the structure functions arise at small x for very large nuclei at
ultra-relativistic energy.  In this description, the effects of
quantum and charge coherence of the partons in the nuclear
wavefunction are properly included.  The gluons arise from the
non-abelian Weizs\"acker-Williams fields generated by the color
charges of the valence quarks.

In this paper, we will extend the treatment of McLerran and
Venugopalan from a description of a single nucleus to the collision of
two nuclei.  This work is in some sense an extension of early effort
which were somewhat ad hoc to describe such collisions by classical
fields \cite{ehtamo,nielson}. We will see that in the region where
most of the parton density sits, the gluon distribution function can
initially be described by a classical field. These classical fields
are to be interpreted as resulting from coherently superimposing large
numbers of gluonic quanta. This way the classical description,
wherever applicable will automatically incorporate coherence effects.

The gluon field for a single nucleus arising in this way is a
non-abelian Weiz\-s\"acker-Williams field.  At the initiation of the
collision, the non-abelian Weiz\-s\"acker-Williams fields of the two
nuclei play the role of boundary conditions for the time evolution of
the gluon field.  This classical field eventually evolves into gluon
quanta.

The picture we have of the collision is therefore the following.
Before the nuclei collide, they are described by valence quarks and
their coherent Weizs\"acker-Williams fields.  These fields are
classical in the sense of classical electromagnetic fields, but of
course can not be thought of as composed of particle with classical
phase space distributions.  During the collision, the fields are still
classical, but sufficiently strong so that the equations of motion
evolve the fields non-linearly with time.  As time evolves, the field
weakens.  When the strength of the gluon field is sufficiently low,
the field equations linearize, and the gluon field describes the
evolution of weakly interacting classical gluon waves.  At this time,
the coherent addition of the fields is no longer important, and they
should be described by an incoherent distribution of gluons.  The
parton cascade model may therefore be used.

Prior to this time however, the coherence in the gluon field is
essential.  The simple fact that the evolution of the gluons is
described by a {\em classical field} is a consequence of the fact that
the gluons are in some locally coherent state. A description in terms
of incoherent {\em classical particles} is simply not possible.

In the second section, we review the relevant results of computation
of the small x structure functions for a single large nucleus.  We
will attempt to describe the kinematic limits of applicability of this
description.  We will argue that the Weizs\"acker-Williams fields
should describe the distribution of gluons in the region of transverse
momenta which gives the dominant contribution after integrating over
transverse momenta.

In the third section, we set up the problem of nucleus-nucleus
scattering.  We derive an equation for the time evolution of the gluon
field.  We relate the results of such a computation to the phase space
density of gluon radiation.

In the fourth section, we summarize our results and speculate on their
region of validity.

\section{Review of the McLerran-Venugopalan Model}

In the work of McLerran and Venugopalan \cite{mv}, it was argued that
for very large nuclei, $A^{1/3} \rightarrow \infty$ at small values of
Bjorken $x$, $x << A^{-1/3}$, the quark and gluon distribution
functions are computable in a weak coupling limit.  This is because
the density of partons per unit area defines a dimensionful scale and
when
\begin{eqnarray}
	\mu^2 \sim {1 \over {\pi R^2}} {{dN} \over {dy}} >> \Lambda_{QCD}^2
\end{eqnarray}
the strong coupling parameter $\alpha_S(\mu^2) $ should become small.
Here $y \sim \ln(1/x)$.

In lowest order in a naive weak coupling expansion, it was shown that
the gluon distribution function was of the Weizs\"acker-Williams form,
that is proportional to $1/x$.  It was also shown that the $p_\perp$
dependence was also of the Weizs\"acker-Williams form $dN/d^2p_\perp
\sim 1/p_\perp^2$ for $\alpha_S \mu << p_\perp << \mu$ where $\mu \sim
\Lambda_{QCD} A^{1/6}$

Of course the naive weak coupling expansion may not be strictly valid,
since there is the well known Lipatov enhancement of the low x
structure functions \cite{lipatov}. This enhancement involves quantum
corrections to the lowest order naive weak coupling result, and
changes the small x distribution to $1/x^{1+C\alpha_s}$.  While this
behavior is computable in the McLerran-Venugopalan model, its nature
is not yet fully understood.  We expect however that as far as the
local effects on the parton distribution at fixed rapidity, $y \sim
\ln(1/x)$, the main effect is to renormalize the charge which
generates the Weizs\"acker-Williams field.

The charge which generates this field in lowest order in the naive
weak coupling expansion is the charge of the valence quarks which are
treated in a no recoil approximation.  While it may be true that the
Lipatov correction might involve new physics, and the picture might
change, we will ignore its effects here except to state that we
believe it will effectively renormalize the valence quark charge
through some x dependent source of charge.  To see how this might
occur, recall that the strength of the Weizs\"acker-Williams
distribution is proportional to the amount of charge present at a
value of x larger than that of the distribution.  We are therefore
assuming the main effect of the quantum fluctuations is to generate an
excess amount of charge at values of x larger than that at which we
measure the parton.

In the work which follow, we will not treat the problems generated by
the Lipatov effect.  We will instead concentrate on the naive lowest
order approximation to the McLerran-Venugopalan model.  This will be
sufficient to understand many qualitative aspects of nucleus-nucleus
collisions, and we hope in the end with small modifications can also
be extended to include the effects of the Lipatov enhancement.

In Refs. \cite{mv}, it was found that to compute the structure
functions one simply treated the valence quarks as a source of light
cone charge.

Here the valence quarks are being treated as a source of charge moving
at the speed of light along the light cone $x^- = 0$.  The source of
charge is being treated classically.  This approximation as justified
so long as the typical transverse momentum scale is
\begin{eqnarray}
	p_\perp << \mu
\end{eqnarray}
where $\mu$ is proportional to the number of valence quarks per unit
area.

At the same time the number of gluon quanta at resolutions with
$p_\perp << \mu$ will be sufficiently high to allow for a description
of the gluonic degrees of freedom through a classical field.

Within these limits, all one has to do is to formulate and solve
the Yang-Mills equations in the presence of the classical current
induced by the valence quarks:
\begin{eqnarray}
\left[D_\mu, F^{\mu \nu}\right] & = &
	U[A](x,z(x))\, J^\nu(z(x))\,  U[A](z(x),x) \qquad .
\end{eqnarray}
Here $z(x) = \left. x\right\vert_{x^+ = 0}$ serves as a reference
point used to define ``initial values'' for the color distribution of
the valence quarks
\begin{eqnarray}
 J^\nu(z(x)) = \delta^{\nu +}
		\delta (x^-) \rho(x_\perp)
\end{eqnarray}
which then, due to covariant current conservation
$\left[D_\nu,J^\nu(x)\right] = 0$ evolve along the particles
trajectory via parallel transport or link operators $ U[A](x,z(x)) :=
{\rm P} \exp -ig \int_{z(x)}^{x}\!\! d{x'}^+ A^- ({x'}^+,x_\perp,x^- =
0) $ connecting the points $x$ and $z(x)$ along the particles
trajectory.

Given a solution to the equations of motion, charge density $\rho$ is
to be treated as a stochastic variable, and to compute ground state
expectation values one must average over all sources with a local
Gaussian weight
\begin{eqnarray}
	\int [d\rho]\  \exp\left\{-\int d^2x_\perp
	{1 \over {2\mu^2}} \rho(x_\perp)^2\right\}
\end{eqnarray}
This Gaussian distribution arose from the approximations used in
Ref. \cite{mv}.  It was argued there that on the transverse resolution
scales corresponding to $p_T << \mu$ that the valence quark charges
may be treated classically.  The exponential factor is the
contribution to the phase space density associated with counting the
number of states of valence charges for a fixed value of the classical
charge.  It can be thought of as arising from the following classical
picture.  Suppose we look in a tube through the nucleus.  This tube
has a transverse size much less than a Fermi but large enough so that
it intersects many nucleons.  In this case, there will typically be
many valence quarks inside the tube each coming from a different
nucleon.  The color charge of each quark is therefore uncorrelated
with that of any other quark and the color charges will add together
in a random walk.  This will lead to the above Gaussian distribution.

Physically, the picture one has is the following: The valence quarks
are recoilless sources of color charge propagating along the light
cone.  Their charge can fluctuate from process to process and the
averaging over charges corresponds to this fluctuation.  The local
charge density is therefore a random variable.  The reason why such a
stochastic source of charge arises is because the transverse
resolution scales which we are interested in are small compared to a
fermi.  On such a scale, when one looks at the nucleus, one sees
uncorrelated quarks coming from different nucleons.  The source of
color charge therefore random walks in color space.  The criteria that
$p_\perp << \mu$ is the criteria that within each transverse
resolution scale, there are many quarks so that the color charge is
typically large and can be treated classically.

The solutions of the above Yang-Mills equations can be chosen to be of
the form
\begin{eqnarray}
	A^+ & = & 0 \nonumber \\
        A^- & = & 0 \nonumber \\
        A^i & = & \theta (x^-) \alpha_i (x_\perp)
\end{eqnarray}
Here the first line may be interpreted gauge choice. Using light cone
gauge $A^+ = 0$ one has direct access to the gluon distribution
functions of the parton model. The requirement to have $A^- =0 $ then
could still be implemented as a gauge choice at least along the
trajectories of the particles, making use of the residual gauge
freedom present in any axial gauge. In this case it turns out that
there is {\em a particular solution} to the equations of motion which
has $A^-$ vanishing everywhere.  On such a solution the link operators
on the right hand side of the Yang-Mills equations drop out entirely
and the equations become
\begin{eqnarray}
	F^{ij} & = & 0 \nonumber \\
        \nabla \cdot \alpha & = & \rho (x_\perp)
\end{eqnarray}
The solution to these equations is that
\begin{eqnarray}
	\alpha_i = -{1 \over{ig}}
		U(x_\perp) \nabla_i U^\dagger (x_\perp)
\end{eqnarray}
that is a pure two dimensional gauge transform of a the vacuum.

Physically, this solution is also easy to understand.  The solution is
a gauge transform of vacuum on one side of the sheet of valence
charge, and another gauge transform of vacuum on the other side of the
sheet.  We have chosen the field to be zero on one side of the sheet
as an overall gauge choice.  (This could be relaxed by an overall
gauge transformation).  Because of the discontinuity in the fields at
the sheet of valence charge, the solution is not a gauge transform of
the vacuum fields.  Its discontinuity gives the source of valence
charge.

Although we have not been successful in explicitly finding the
solution to this equation, it is in principle possible to do
numerically.  Several generic features of the averaging over different
sources of charge are possible to infer nevertheless.

For $p_\perp \le \alpha_S \mu$, the typical value of the external
charge is so large that it is a bad approximation to linearize the
gauge transformation and directly compute the field in a naive weak
coupling expansion.  In this region, the non-linearities of the field
equation become important.

\begin{figure}[htb]
	\begin{minipage}[b]{\medfigsize} \epsfxsize \medfigsize
		\epsfbox{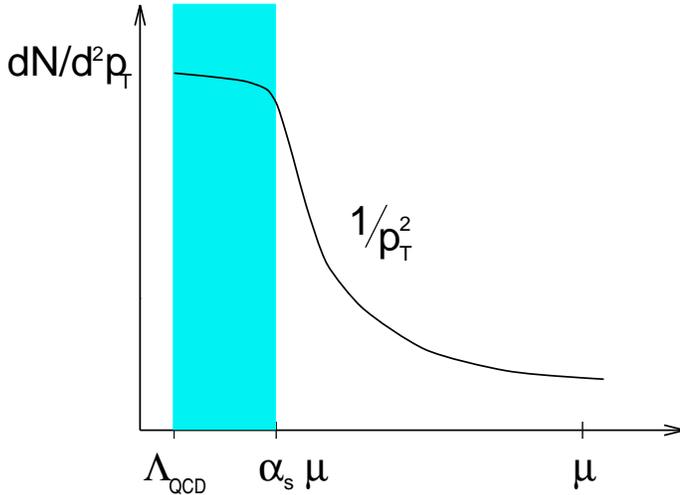}
	\end{minipage} \hfill
	\begin{minipage}[b]{\medcapsize} \caption[] {
		\label{distribution} \sloppy \small
	 Weizs\"acker-Williams distribution for a single nucleus
	 }
	\vspace{5 truecm} \end{minipage}
\end{figure}

In this kinematic region, the shape of the Weizs\"acker-Williams
distribution changes form, as is shown in Fig. \ref{distribution}.

The $1/p^2_\perp$ behavior turns over and goes to a constant.  This
provides a low momentum cutoff in the number of gluons generated by
the distribution.  At high momentum, we can compute no further than
$p_\perp \le \mu$.  The distribution should nevertheless extend beyond
this region.  In fact the upper momentum cutoff should be determined
only by the kinematic limit of the process considered. Strictly
speaking the number of Weizs\"acker-Williams gluons is infinite, but
only logarithmically, and the cutoff will be determined by the process
of physical interest.

Suppose we consider the production of gluons in a two nucleus
collision.  We can naively determine the cutoff in momentum of
produced gluons.  Lets us assume that we are in the weak coupled
perturbative region.  The rate must be proportional to the density
squared of gluons, so that it must be proportional to
$\alpha_S^2\mu^4$.  It involves scattering, so their are two more
factors of $\alpha_S$.  Therefore on dimensional grounds, we expect
that the probability of making a pair of on mass shell gluons should
be of order $\alpha_S^4 \mu^4 /p_\perp^4$.  This is of order one when
$p_\perp \sim \alpha \mu$, that is at precisely the place where the
fields evolve non-linearly.  It cuts off rapidly in $p_\perp$, so that
the number of produced gluons should be ultraviolet finite.

This example teaches us two things: First that for a physical process
there is no ultraviolet divergence and for this process the important
contribution for gluon production is at scales less than $\mu$.
Second, that the process is strongest in the region where the field is
strong.  In this region, the field is evolving non-linearly, and the
coherence of the field is important.  It would therefore be a mistake
to assume that the distribution of produced gluons reflects the
distribution in the initial nuclei.  This is true for gluons with
$p_\perp \ge \alpha_S \mu$, that is 'hard gluons', but the softer
gluons which dominate the production are in a non-linear region.

The appearance of these non-linearities might be qualitatively
included in the parton cascade model.  However insofar as there is an
infrared cutoff dependence in some physical process, the results will
be somewhat quantitatively unreliable.  Processes without such a
cutoff dependence would of course be more reliably computed.

The hope will be in our attempt to compute gluon production is that
the classical non-linearities will cutoff the the naive divergence in
the production amplitude at small $p_\perp$.  To see that this is
plausible, recall that the single gluon distribution changes its from
at small $p_\perp$ from $1/p_\perp^2$ to constant at $p_\perp \le
\alpha_S \mu$.  It is therefore quite plausible that these effects in
fact cutoff the singularity at some scale of order $\alpha_S \mu$.  If
this is so, then if $A$ is large enough so that $A^{1/6} >> 1$, this
cutoff is at a scale much larger than $\Lambda_{QCD}$, and the
computation is self-consistent.

\section{The Two Nucleus Problem}

We now turn to the problem of nucleus-nucleus scattering.  We work in
the center of mass frame.  Both nuclei are at sufficiently high energy
so that they can be treated as infinitesimally thin sheets.  They are
large enough so that these sheets can be taken to be of infinite
extent in the transverse direction.  We will be interested in
describing the production of gluons at typical momentum scales which
are $p_\perp << \mu$, but much larger than $200~MeV$.  In this case
the source of color charge can be taken as classical as in the
McLerran-Venugopalan model.  The charge is of course a stochastic
variable which must be integrated over with a Gaussian weight as
described in the previous section.

\begin{figure}[htb]
	\begin{minipage}[b]{\medfigsize} \epsfxsize \medfigsize
		\epsfbox{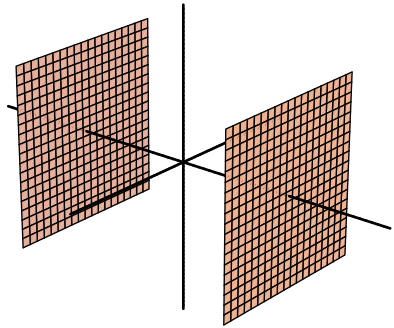}
	\end{minipage} \hfill
	\begin{minipage}[b]{\medcapsize}
	\caption[] {
		\label{twonuclei} \sloppy \small Two nuclei
		Lorentz-contracted to infinitely
		thin sheets before
		the collision takes place }
	\vspace{3 truecm}
	\end{minipage}
\end{figure}
The sources of color field set up a classical color field.  After the
collision the color field will begin to evolve in time.  Much after
the collision, the color field will describe the propagation of free
gluons.  In this section we will describe how to compute the evolution
of the color field, and then how to compute the final state
distribution of gluons.

The Yang-Mills equation for the two source problem is
\begin{eqnarray}
	\left[D_\mu, F^{\mu \nu}\right] = J^\nu (x)
	\label{YMcoll}
\end{eqnarray}
where
\begin{eqnarray}
	J^+ & = & \delta (x^-) \rho_1 (x_\perp)\nonumber \\
        J^- & = & \delta (x^+) \rho_2 (x_\perp) \nonumber \\
        J^i & = & 0
\end{eqnarray}
and we have restricted ourselves to work in a gauge where the link
operators along the particle trajectories drop out.

\begin{figure}[htb]
	\begin{minipage}[b]{\medfigsize} \epsfxsize \medfigsize
		\epsfbox{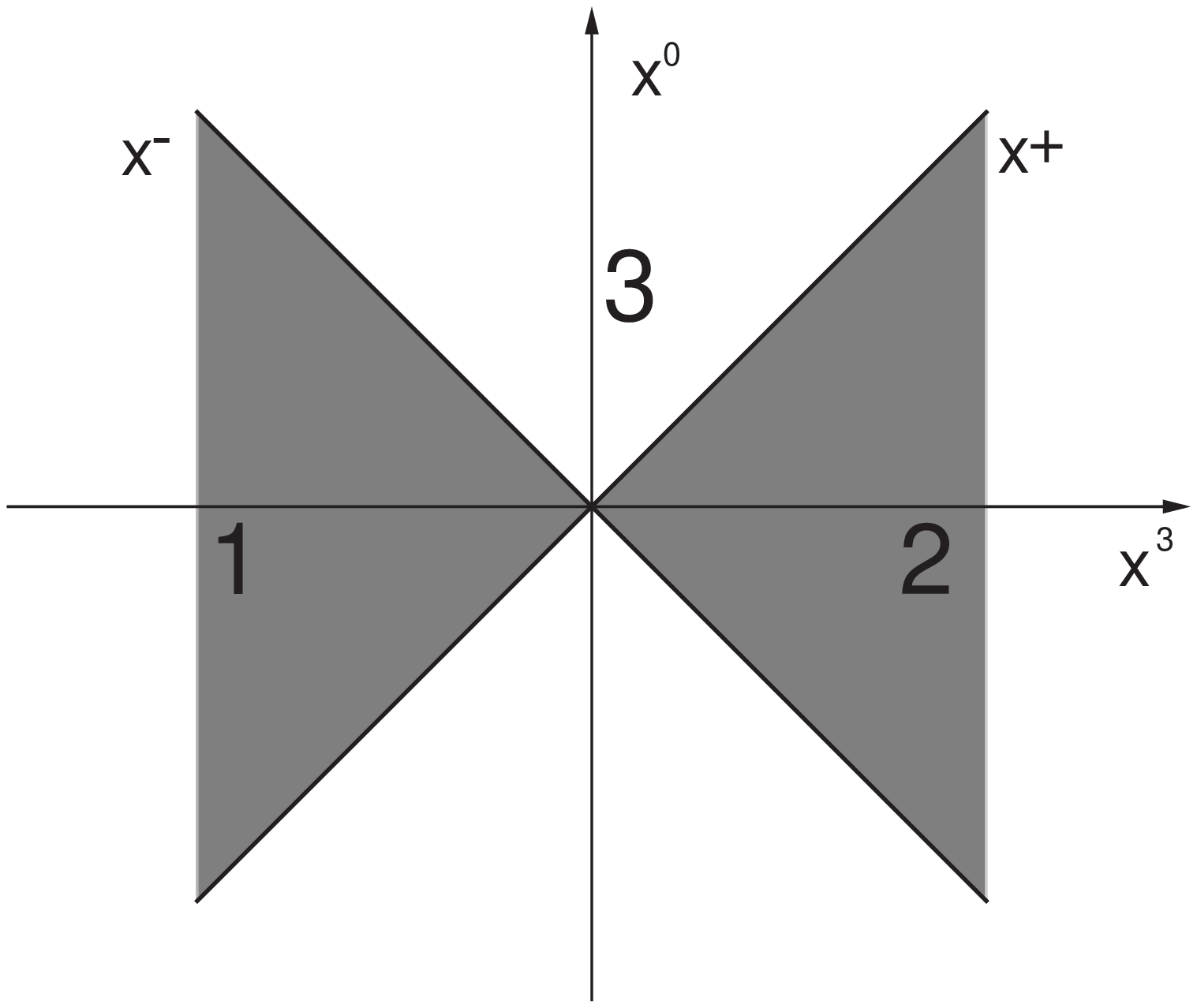}
	\end{minipage} \hfill
	\begin{minipage}[b]{\medcapsize} \caption[] {
		\label{regions} \sloppy \small Regions with different
		structures of the gauge potential: \\
	       	In regions 1 and 2 we have the well known one nucleus
		solutions $\alpha_{1,2}$. While in the backward light
		cone there the gauge potential is vanishing we have a
		nontrivial solution in the forward lightcone, region 3
		}
	\vspace{1 truecm} \end{minipage}
\end{figure}

Before the collision takes place, we find a solution of the equations
of motion to be
\begin{eqnarray}
	A^+ & = & 0 \nonumber \\
        A^- & = & 0 \nonumber \\
        A^i & = & \theta (x^-) \theta(-x^+) \alpha^i_1 (x_\perp) +
 \theta (x^+) \theta (-x^-) \alpha^i_2 (x_\perp)
\end{eqnarray}
This is a solution of the Yang-Mills equations in all of space-time
except on or within the forward light cone, as shown in
Fig. \ref{regions}.
In the forward light cone, we must add in extra pieces in order to
have a solution.  This will be done below.  The two dimensional vector
potentials are pure gauges and solve for $t < 0$
\begin{eqnarray}
	\nabla \cdot \alpha_{1,2} =  \rho_{1,2}(x_\perp)
\end{eqnarray}

The physical picture one has of this solution prior to the collision
is that the nuclei have zero field in front of them as they approach
one another.  Behind them the field is pure gauge.  Because each
nucleus has a different charge density, the gauge is different for
each nucleus.  This is an exact solution of the equations of motion so
long as one is outside the forward light cone, that is in regions
which are out of causal contact from the collision event.

The fact that we have a solution of the equations of motion which does
not evolve in time before the collision is remarkable.  This solves
the problem of cascade models that an isolated nucleus composed of
partons will spontaneously fall apart.  Here the individual nuclei and
their parton clouds are static except for their overall center of mass
motion.

How do we describe the fields after the collision?  Except at the
forward light cone, that is, when we are inside or outside the cone,
the fields satisfy free field equations.  We will look for a Lorentz
covariant solution to the equations of motion.  We try, for $x^- > 0$
and $x^+ > 0$
\begin{eqnarray}
	A^+ & = & x^+ \alpha(\tau, x_\perp) \nonumber \\
        A^- & = & x^- \beta(\tau, x_\perp) \nonumber \\
        A^i & = & \alpha_3^i (\tau,x_\perp)
\end{eqnarray}
where
\begin{eqnarray}
	\tau = \sqrt{t^2 - z^2} = \sqrt{2x^+x^-}
\end{eqnarray}
This solution only depends on the longitudinal boost invariant
variable $\tau$ and has no dependence on the space-time rapidity
variable
\begin{eqnarray}
	\eta = { 1 \over 2} \ln{x^+ \over x^-}
\end{eqnarray}
The factors of $x^+$ and $x^-$ in the definition of the vector
potential guarantee that under longitudinal boosts, the vector
potential transforms properly.

By making a gauge transformation which is only a function of proper
time and $x_\perp$,
\begin{eqnarray}
	U = U(\tau, x_\perp)
\end{eqnarray}
we see that we can fix
\begin{eqnarray}
	\beta (\tau, x_\perp) = -\alpha(\tau,x_\perp)
\end{eqnarray}
which we shall choose to do. This choice corresponds to a gauge
condition
\begin{eqnarray}
x^+ A^- + x^- A^+ & = & 0
\end{eqnarray}
which in turn is consistent with dropping the link-operators in
(\ref{YMcoll}).

If such a solution solves the equations of motion an boundary
conditions, it will predict that the distribution of partons is boost
invariant.  It is the generalization therefore of Bjorken's boost
invariant hydrodynamic equations to the equations which generate the
initial conditions for the hydrodynamic equations.

As a consequence of the boost invariance of the Yang-Mills equations,
the ansatz above solves the equations within the forward light cone.
This can be checked explicitly. The equations which result for
$\alpha_3^i$, and $\alpha$, for $x^+, x^- > 0$ are
\begin{eqnarray}
{1\over\tau^3}\partial_\tau \tau^3 \partial_\tau\alpha
	- \left[ D^i, \left[D^i,\alpha\right] \right]
	& = & 0
\nonumber \\
{1\over\tau} \left[ D_i,
	\partial_\tau \alpha_3^i \right]  +
	ig \tau \left[\alpha,\partial_\tau\alpha\right] & = & 0
\nonumber \\
{1\over \tau}\partial_\tau \tau \partial_\tau\alpha^i_3
	- ig \tau^2\left[
	\alpha, \left[D^i,\alpha\right]
	\right] - \left[D^j,F^{ji}\right] & = & 0
\end{eqnarray}

These four equations can be checked to be consistent with on another.
{From} this point on, all vector indices will refer to two dimensional
transverse vectors.  The longitudinal and time coordinates will be
denoted separately.

In intermediate step in deriving the above equations, we computed the
field strengths $F^{\mu \nu}$.  The results of that computations are
\begin{eqnarray}
F^{+-} & = &  -{1\over \tau}\partial_\tau \tau^2 \alpha
\nonumber \\
F^{ij} & = & \partial^i \alpha_3^j - \partial^j \alpha_3^i
	-ig \left[\alpha^i_3,\alpha^j_3\right]
\nonumber \\
F^{i\pm} & = &
	-x^\pm \left( {1\over \tau}
		\partial_\tau \alpha^i_3 \mp
		 \left[D^i,\alpha\right]\right)
\end{eqnarray}

The only task which remains is to show that the above solution also
satisfies the boundary conditions generated by the sources.  For
either $x^- < 0$ or $x^+ < 0$, the ansatz above satisfies the
equations trivially.  We look first at the equation $\left[D_\mu,
F^{\mu i}\right] = 0$.  This has a delta function singularity at $x^-
= x^+ = 0$ which requires that
\begin{eqnarray}
	\left.\alpha_3^i(0,x_\perp)\right\vert_{\tau=0}
		= \alpha_1^i (x_\perp)+\alpha_2^i (x_\perp)
\end{eqnarray}
and there are no further discontinuities in this equation.

Now for $\left[D_\mu, F^{\mu \pm}\right] = J^{\pm}$, we find that
\begin{eqnarray}
	\left.\alpha^{\vphantom{i}}
		(\tau,x_\perp)\right\vert_{\tau=0} = -{ig \over 2}
	\left[{\alpha_1}_i(x_\perp), \alpha^i_2(x_\perp)\right]
\end{eqnarray}
These two equations for $J^\pm$
reduce to the same boundary condition, and
therefore neither $\alpha$ nor $\alpha_3^i$ are overconstrained,
demonstrating once more that our ansatz contains the correct degrees
of freedom.

Note that assuming the boundary conditions above, we are implicitly
requiring that the solution be regular at $\tau = 0$.  It is easy to
check that the quantities $\alpha$ and $\alpha^i_3$ can either be
regular at the origin or diverge like $\alpha \sim 1/\tau^2$ and
$\alpha^i \sim \ln(\tau)$.  These singular solutions will lead to a
divergent energy density, and are therefore not allowed.

So with the above two boundary conditions, the solution to the
equations of motion are uniquely specified.  This solution is
remarkable since in spite of the possible asymmetry in the charge on
either nuclei, the solution is up to trivial factors rapidity
independent.  This has amusing phenomenological consequences for the
collisions of asymmetric nuclei.  The distribution should be flat in
rapidity.  The height of the central plateau of course depends on the
asymmetry in a non-trivial way.

In order to determine the gluon radiation produced by these fields, we
must solve them at proper times long after the collision.  We expect
that the energy density will dissipate and therefore the field
strengths will become small.  Using the expressions above, we conclude
that asymptotically, for large $\tau$,
\begin{eqnarray}
	\alpha(\tau, x_\perp) & \rightarrow &
	V(x_\perp) \epsilon(\tau, x_\perp) V^\dagger
	(x_\perp) \nonumber \\
        \alpha^i_3(\tau, x_\perp) & \rightarrow &
	V(x_\perp)\left[ \epsilon^i(\tau, x_\perp)
	 -{1 \over {ig}} \partial^i\right]
	V^\dagger (x_\perp)
\end{eqnarray}
The solution should tend to a small field plus a gauge transformation.
The value of this gauge transformation is determined by the field
equations and has a non-trivial dependence on the sources.  It results
from solving the non-linear time evolution equations for the fields.

The equations of motion for the fields in the asymptotic region are
linear for $\epsilon$ and $\epsilon^i$.  The equations are
\begin{eqnarray}
	{1 \over \tau^3} \partial_\tau \tau^3 \partial_\tau \epsilon
 -\nabla^2 \epsilon & = & 0 \nonumber \\
     {1 \over \tau} \partial_\tau \tau \partial_\tau \epsilon^i
       -\left(\nabla^2 \delta^{ij}
		- \nabla^i\nabla^j\right)\epsilon^j
	& = & 0
\end{eqnarray}
Observe, that $\nabla^i \epsilon^i$ does not enter the asymptotic
equations, so that there are in fact only two dynamical degrees of
freedom in the solution, as must be the case.

The solutions to the above equations at asymptotically large $\tau$
are of the form
\begin{eqnarray}
	\alpha^a(\tau,x_\perp) & = &
      \int {{d^2k_\perp} \over {(2\pi)^2}}
       {1 \over \sqrt{2\omega}}
      \left\{ a_1^a(\vec{k}_\perp) {1 \over \tau^{3/2}}
      e^{ik_\perp\cdot x_\perp -i\omega \tau} + C. C.
        \right\} \nonumber \\
        \vec{\alpha}^{a,i} (\tau,x_\perp) & = &
	\int {{d^2k_\perp} \over {(2\pi)^2}}
        \kappa^i
{1 \over \sqrt{2\omega}} \left\{ a_2^a(k_\perp)
{1 \over \tau^{1/2}}
      e^{ik_\perp x_\perp-i\omega \tau} + C. C. \right\}
\end{eqnarray}
In this equation, the frequency $\omega = \mid k_\perp \mid$, and the
vector
\begin{eqnarray}
	\kappa^i = \epsilon^{ij} k^j/\omega
\end{eqnarray}
The notation $+ C. C.$ means to add in the complex conjugate piece.

To derive an expression for the energy density, we recall that $\tau $
is large.  Near $z = 0$, this implies that the range of $z$ where
$\tau \sim t >> z$ that the solutions are $z$ independent.  This means
they asymptotically have zero $p_z$.  Now suppose we are at any value
of $z$, and $\tau$ is large but $t \sim z$.  We can do a longitudinal
boost to $z = 0$ without changing the solution.  Again in this frame
the solution has zero $p_z$.  We see therefore that for the asymptotic
solutions that the space time rapidity is one to one correlated with
the momentum space rapidity, that is at asymptotic times we find that
\begin{eqnarray}
	\eta = {1 \over 2} \ln(x^+/x^-)
	= y = {1 \over 2} \ln(p^+/p^-)
\end{eqnarray}

To proceed further, we compute the energy density in the neighborhood
of $z = 0$.  Here asymptotically $\tau = t$.  The energy in a box of
size $R$ in the transverse direction and $dz$ in the longitudinal
direction, with $L << t$ becomes \cite{blaizot}
\begin{eqnarray}
	dE = {{dz} \over t} \int {{d^2k_\perp}\over {(2\pi)^2}}
	\omega \sum_{i,b} \mid a_i^b(k_\perp) \mid^2
\end{eqnarray}
Recalling that $dy = dz/t$, we find that
\begin{eqnarray} {{dE} \over {dyd^2k_\perp}} =
{1 \over {(2\pi)^3}} \omega \sum_{i,b} \mid a_i^b(k_\perp) \mid^2
\end{eqnarray}
and the multiplicity distribution of gluons is
\begin{eqnarray} {{dN} \over
{dyd^2k_\perp}} = {1 \over \omega} {{dE} \over {dyd^2k_\perp}}
\end{eqnarray}
As we expect for a boost covariant solution, the multiplicity
distribution is rapidity invariant.

Finally, we must comment a bit on the characteristic time scale for
the dissipation of the non-linearities in the equations for the time
dependent Weizs\"acker-Williams fields.  This is difficult to estimate
in general, but scaling arguments should suffice to estimate the time
scale.  The basic point is that the typical momentum scale in the
problem relevant for the formation of most of the gluons is $p_\perp
\sim \alpha \mu$ which up to logarithms is $p_\perp \sim 200 MeV
A^{1/6}$.  The characteristic time scale for the dissipation of the
classical non-linearities should therefore be of order $\tau \sim 1
/p_\perp$.  This is in agreement with other estimate of the
characteristic formation time for partons, and is in agreement with
the model of Geiger and M\"uller.

\section{Summary and Conclusions}

We have derived a theory of the formation of gluons which is
applicable for small x gluons in the collisions of very large nuclei.
We have found that the gluon distributions as measured in deep
inelastic scattering undergo an entirely non-trivial evolution in
forming gluons which would be the initial conditions for a parton
cascade.  We have shown how one can compute these initial conditions.

After finding the initial conditions, the subsequent evolution might
be described by a combination of the parton cascade and hydrodynamics.

There are many further problems to be addressed in this theory. The
equations described above must be numerically solved. This will
provide for initial conditions for average head on collisions.  In
addition it will predict the spectrum of fluctuations from collisions
to collision.  Perhaps the most interesting problem is to compute the
hard particles produced during the early evolution of the
distributions so as to find a precise quantitative test of the theory.

\section*{Acknowledgments}

This research was supported by the U.S. Department of Energy under
grants No. DOE High Energy DE--FG02--94ER40823 and No. DOE Nuclear
DE--FG02--87ER--40328. One of us (HW) was supported by a fellowship of
the Alexander von Humboldt Foundation.  Larry McLerran wishes to thank
Klaus Geiger, Berndt M\"uller and Raju Venugopalan who carefully read
an early version of this manuscript and provided him with useful
suggestions.  He also wishes to thank Judah Eisenberg, Emil Mottola,
and Ben Svetitsky for organizing a workshop at the International
Center for Theoretical Nuclear Physics in Trento where these ideas
were initiated, and the hospitality of the staff who provided such a
productive work environment.  He also thanks Miklos Gyulassy who at
this meeting provided many insightful suggestions.

\end{document}